\documentclass[aps,epsf,preprint]{revtex4}

\renewcommand{\slash}[1]{#1 \hspace{-0.55em} / }

\def\bq{\begin{eqnarray}}
\def\eq{\end{eqnarray}}
\def\be{\begin{eqnarray}}
\def\ee{\end{eqnarray}}
\def\ben{\begin{enumerate}}\def\een{\end{enumerate}}

\def\roughly#1{\mathrel{\raise.3ex\hbox{$#1$\kern-.75em
\lower1ex\hbox{$\sim$}}}}

\tighten

\begin{document}


\def\bra{\langle }
\def\ket{\rangle }

\title{Conventional nuclear effects on
generalized parton distributions
of trinucleons
}

\author{
S. Scopetta}
\address{
Dipartimento di Fisica, Universit\`a degli Studi
di Perugia, via A. Pascoli
06100 Perugia, Italy
\\
and INFN, sezione di Perugia}

\begin{abstract}

The measurement of nuclear Generalized Parton Distributions
(GPDs) will represent a valuable tool to understand the structure
of bound nucleons in the nuclear medium, as well as the role
of non-nucleonic degrees of freedom
in the phenomenology of hard scattering off
nuclei. By using a realistic microscopic approach for the evaluation
of GPDs of $^3He$, it will be shown that conventional nuclear
effects, such as isospin and binding ones,
or the uncertainty related to the use of a given 
nucleon-nucleon potential,
are rather bigger than 
in the forward case. These findings suggest that, 
if great attention is not paid
to infer the properties of nuclear GPDs 
from those of nuclear parton distributions,
conventional nuclear effects 
can be easily mistaken for exotic ones.
It is stressed therefore that  
$^3He$, for which the best realistic calculations are possible,
represents a unique target to discriminate between conventional and exotic
effects. The complementary information which could be obtained
by using a $^3H$ target is also addressed.

\end{abstract}

\maketitle
\section{Introduction}

The measurement of
Generalized Parton Distributions (GPDs) \cite{first}, 
parametrizing  the non-perturbative hadron structure
in hard exclusive 
processes,
represents one of the challenges of nowadays hadronic Physics
(for reviews, 
see, e.g.,  \cite{jig,pog,dpr,rag,bp}).
GPDs enter
the long-distance dominated part of
exclusive lepton Deep Inelastic Scattering
(DIS) off hadrons.
Deeply Virtual Compton Scattering (DVCS),
i.e. the process
$
e H \longrightarrow e' H' \gamma
$ when
$Q^2 \gg m_H^2$,
is one of the the most promising to access GPDs
(here and in the following,
$Q^2$ is the momentum transfer between the leptons $e$ and $e'$,
and $\Delta^2$ the one between the hadrons $H$ and $H'$)
\cite{first,gui}.
Relevant experimental efforts to measure GPDs
are taking place, and
a few DVCS data have been already published 
\cite{hermes,clas}. 
The issue of measuring GPDs for nuclei
has been addressed in several papers. In the first one
\cite{cano1}, 
it was shown that
the knowledge of GPDs would permit the investigation
of the short light-like distance structure of nuclei, and thus the interplay
of nucleon and parton degrees of freedom in the nuclear 
wave function.
In inclusive DIS off a nucleus
with four-momentum $P_A$ and $A$ nucleons of mass $M$,
this information can be accessed in the 
region where
$A x_{Bj} \simeq {Q^2\over 2 M \nu}>1$,
being $x_{Bj}= Q^2/ ( 2 P_A \cdot q )$ and $\nu$
the energy transfer in the laboratory system.
In this region measurements are very difficult, because of 
vanishing cross-sections. As explained in \cite{cano1}, 
the same physics can be accessed
in DVCS at much lower values of $x_{Bj}$.
In Ref. \cite{poly} it has been shown that, for finite nuclei,
the measurement of GPDs would provide
us with peculiar information
about the spatial distribution of energy, momentum
and forces experienced by quarks and gluons inside hadrons.
This argument has been retaken and confirmed recently in Ref. \cite{guz1}.
DVCS has been extensively discussed for different
nuclear targets.
Impulse Approximation (IA) calculations,
supposed to give the bulk of nuclear
effects at $0.05 \leq A x_{Bj} \leq 0.7$,
have been performed for
the deuteron \cite{cano2}
and for spinless nuclei
\cite{gust}, in particular for $^4He$ \cite{liuti},
for which an experiment is going on at JLab \cite{4he}.  
For nuclei of any spin, estimates of GPDs have been
provided and prescriptions for
nuclear effects have been proposed in \cite{km}.
Analyses of nuclear DVCS beyond IA,
with estimates of shadowing
effects and involving therefore large light-like distances and correlations
in nuclei have been also performed \cite{frest,guz2}. 
While several studies have shown that the measurement
of nuclear GPDs can unveil crucial information on possible
medium modifications of nucleons in nuclei
\cite{liuti2,thomas}, great attention has to be paid to avoid
to mistake them with
conventional nuclear effects.
To this respect, a special role
would be played by few body nuclear targets,
for which realistic studies 
are possible and exotic effects, such as
the ones of non-nucleonic degrees of freedom,
not included in a
realistic wave function,
can be disentangled.
To this aim,
in Ref. \cite{io}, a realistic IA calculation
of the quark unpolarized GPD $H_q^3$ of
$ ^3He $ has been presented.
The study of GPDs for $^3He$ is interesting
for many aspects. 
In fact, $^3He$ is a well known nucleus, 
and it is extensively used as an effective neutron target:
the properties of the free neutron are being investigated
through experiments with nuclei
(the measurement of neutron GPDs
using nuclei has been discussed in Ref. \cite{guz3}), whose data
are analyzed taking nuclear effects properly into account.
For example, it has been shown, firstly in \cite{thom1},
that unpolarized DIS off trinucleons ($^3H$ and $^3He$) can provide 
relevant information on PDFs at large $x_{Bj}$, 
while it is known since a long time that 
its particular spin structure
suggests the use of $^3He$ as an effective polarized
neutron target \cite{friar,io2,sauer,gfst}.
Polarized $^3He$ will be therefore the first candidate
for experiments aimed at the study of spin-dependent
GPDs of the free neutron. 
In Ref. \cite{io}, the GPD $H_q^3$ of
$ ^3He $ has been evaluated
using a realistic non-diagonal spectral function,
so that momentum and binding effects are rigorously estimated.
The scheme proposed in that paper is valid for $\Delta^2 \ll Q^2,M^2$
and it permits to calculate GPDs in the kinematical range relevant to
the coherent, no break-up channel of deep exclusive processes off $^3He$.
In fact, the latter channel can be hardly studied at
large $\Delta^2$, due to the vanishing cross section
\cite{frest}.
Nuclear effects are found to be larger than in the forward case
and to increase with $\Delta^2$ at fixed
skewedness, and with the skewedness at fixed $\Delta^2$.
In particular the latter $\Delta^2$ dependence
does not simply factorize,
in agreement with previous findings for the deuteron target
\cite{cano2} and at variance
with prescriptions proposed for finite nuclei \cite{km}. 

Here, the analysis of Ref. \cite{io} is 
extended into various directions.
The main point of the paper will be to
stress that the properties of nuclear GPDs should not be trivially
inferred from those of nuclear parton distributions. 
After a fast summary of the formalism of Ref. \cite{io},
in the third section of the paper
a detailed study of the flavor dependence of the nuclear
effects, which is due to the fact that $^3He$ is non isoscalar,
is carried on; a serious warning concerning the
possibility to use momentum distributions instead of
spectral functions is motivated; the dependence on
the choice of the nucleon-nucleon potential used
to estimate the nuclear GPDs is shown.
In the following section, the information which could be obtained
by using a $^3H$ target is addressed.
Eventually, conclusions are drawn in the fifth section. 

\section{Formalism}

The definitions of GPDs of
Ref. \cite{jig} is used.
For a spin $1/2$ hadron target, with initial (final)
momentum and helicity $P(P')$ and $s(s')$, 
respectively, 
the GPDs $H_q(x,\xi,\Delta^2)$ and
$E_q(x,\xi,\Delta^2)$
are defined through the light cone correlator
\begin{eqnarray}
\label{eq1}
F^q_{s's}(x,\xi,\Delta^2) & = &
{1 \over 2} \int {d \lambda \over 2 \pi} e^{i \lambda x}
\bra P' s' | \, \bar \psi_q \left(- {\lambda n \over 2}\right)
\slash{n} \, \psi_q \left({\lambda n \over 2} \right) | P s \ket  =  
\nonumber
\\
& = & H_q(x,\xi,\Delta^2) {1 \over 2 }\bar U(P',s') 
\slash{n} U(P,s) + 
E_q(x,\xi,\Delta^2) {1 \over 2} \bar U(P',s') 
{i \sigma^{\mu \nu} n_\mu \Delta_\nu \over 2M} U(P,s)~,
\end{eqnarray}
\\
where 
$\Delta=P^\prime -P$
is the 4-momentum transfer to the hadron,
$\psi_q$ is the quark field and M is the hadron mass.
It is convenient to work in
a system of coordinates where
the photon 4-momentum, $q^\mu=(q_0,\vec q)$, and $\bar P=(P+P')/2$ 
are collinear along $z$.
The skewedness variable, $\xi$, is defined as
\bq
\xi = - {n \cdot \Delta \over 2} = - {\Delta^+ \over 2 \bar P^+}
= { x_{Bj} \over 2 - x_{Bj} } + {\cal{O}} \left ( {\Delta^2 \over Q^2}
\right ) ~,
\label{xidef}
\eq
where $n$
is a light-like 4-vector
satisfying the condition $n \cdot \bar P = 1$.
(Here and in the following, $a^{\pm}=(a^0 \pm a^3)/\sqrt{2}$).
In addition to the variables
$x,\xi$ and $\Delta^2$, GPDs depend,
on the momentum scale $Q^2$. 
Such a dependence, not discussed here, will be 
omitted.
The constraints of $H_q(x,\xi,\Delta^2)$ are: 

i) the 
``forward'' limit, 
$P^\prime=P$, i.e., $\Delta^2=\xi=0$, yielding the usual PDFs
\bq
H_q(x,0,0)=q(x)~;
\label{i)}
\eq

ii)
the integration over $x$, yielding the contribution
of the quark of flavour $q$ to the Dirac 
form factor (f.f.) of the target:
\bq
\int dx H_q(x,\xi,\Delta^2) = F_1^q(\Delta^2)~;
\label{ii)}
\eq

iii) the polynomiality property \cite{jig},
involving higher moments of GPDs, according to which
the $x$-integrals of $x^nH^q$ and of $x^nE^q$
are polynomials in $\xi$ of order $n+1$.

In Ref. \cite{epj},
an expression
for $H_q(x,\xi,\Delta^2)$ of a given hadron target, 
for small values of $\xi^2$, has been obtained
from the definition Eq. (\ref{eq1}).
The approach has been later applied 
in Ref. \cite{io} to obtain the GPD 
$H_q^3$ of $^3He$ in IA, as
a convolution between
the non-diagonal spectral function of the internal nucleons,
and the GPD $H_q^N$ of the nucleons themselves. 
Let me recall the main formalism of Ref. \cite{io},
which will be used in this paper.
In the class of frames discussed above,
and in addition to the kinematical variables
$x$ and $\xi$, already defined,
one needs the corresponding ones for the nucleons in the target nuclei,
$x'$ and $\xi'$. 
The latter quantities can be obtained defining the ``+''
components of the momentum $k$ and $k + \Delta$ of the struck parton
before and after the interaction, with respect to
$\bar P^+$ and $\bar p^+ = {1 \over 2} (p + p')^+$:
\begin{eqnarray}
k^+ & = & (x + \xi ) \bar P^+ =  (x' + \xi') \bar p^+~,
\\
(k+\Delta)^+ & = & 
(x - \xi ) \bar P^+ =  (x' - \xi') \bar p^+~,
\end{eqnarray}
so that
\begin{eqnarray}
\xi' & = & - { \Delta^+ \over 2 \bar p^+}~,
\label{defxi1}
\\
x' & = & {\xi' \over \xi} x~,
\label{defx1}
\end{eqnarray}
and, since $\xi = - \Delta^+ / (2 \bar P^+)$, 
if $\tilde z = p^+/P^+$
\begin{eqnarray}
\xi ' = {\xi \over \tilde z( 1 + \xi ) - \xi}~.
\label{xi1}
\end{eqnarray}

In Ref. \cite{io},
a convolution formula for $H_q^3$ has been derived in IA,
using the standard procedure developed in studies of
DIS off nuclei \cite{fs,cio,ia}.
It reads:
\begin{eqnarray}
H_q^3(x,\xi,\Delta^2) & \simeq & 
\sum_N \int dE \int d \vec p
\, 
[ P_{N}^3(\vec p, \vec p + \vec \Delta, E ) + 
{\cal{O}} 
( {\vec p^2 / M^2},{\vec \Delta^2 / M^2}) ]
\nonumber
\\
& \times & 
{\xi' \over \xi}
H_{q}^N(x',\xi',\Delta^2) + 
{\cal{O}} 
\left ( \xi^2 \right )~.
\label{spec}
\end{eqnarray}
In the above equation, 
$P_{N}^3 (\vec p, \vec p + \vec \Delta, E )$ is
the one-body non-diagonal spectral function
for the nucleon $N$,
with initial and final momenta
$\vec p$ and $\vec p + \vec \Delta$, respectively,
in $^3He$:
\begin{eqnarray}
P_N^3(\vec p, \vec p + \vec \Delta, E)  & = & 
{1 \over (2 \pi)^3} {1 \over 2} \sum_M 
\sum_{R,s}
\bra \vec P'M | (\vec P - \vec p) S_R, (\vec p + \vec \Delta) s\ket 
\bra (\vec P - \vec p) S_R,  \vec p s| \vec P M \ket
\times
\nonumber
\\
& \times &  
\, \delta(E - E_{min} - E^*_R)~,
\label{spectral}
\end{eqnarray}
and the quantity $H_q^N(x',\xi',\Delta^2)$
is
the GPD of the bound nucleon N
up to terms of order $O(\xi^2)$ (note that in its
definition 
use has been made
of Eqs. (\ref{defxi1}) and (\ref{defx1})).

The delta function in Eq (\ref{spectral})
defines $E$, the removal energy, in terms of
$E_{min}=| E_{^3He}| - | E_{^2H}| = 5.5$ MeV and
$E^*_R$, the excitation energy 
of the two-body recoiling system.
The main quantity appearing in the definition
Eq. (\ref{spectral}) is
the overlap integral
\bq
\bra \vec P M | \vec P_R S_R, \vec p s \ket=
\int d \vec y \, e^{i \vec p \cdot \vec y}
\bra \chi^{s},
\Psi_R^{S_R}(\vec x) | \Psi_3^M(\vec x, \vec y) \ket~,
\label{trueover}
\eq 
between the eigenfunction 
$\Psi_3^M$ 
of the ground state
of $^3He$, with eigenvalue $E_{^3He}$ and third component of
the total angular momentum $M$, and the
eigenfunction $\Psi_R^{S_R}$, with eigenvalue
$E = E_{min}+E_R^*$ of the state $R$ of the intrinsic
Hamiltonian pertaining to the system of two interacting
nucleons \cite{over}.
As discussed in Ref. \cite{io}, the accuracy of the calculations
which will be presented, since a NR spectral function
will be used to evaluate Eq. (\ref{spec}), is
of order 
${\cal{O}} 
\left ( {\vec p^2 / M^2},{\vec \Delta^2 / M^2} \right )$,
or, which is the same,
$\vec p^2, \vec \Delta^2 << M^2$.
While the first of these conditions is the usual one
for the NR treatment of nuclei, the second
forces one to use Eq. (\ref{spec}) only
at low values of $\vec \Delta^2$, for which the accuracy is good enough.
The interest of the present
calculation is indeed to investigate nuclear effects at low values
of $\vec \Delta^2$, for which measurements in the coherent channel 
may be performed.
The main emphasis of the present approach, as already said,
is not on the absolute values of the results, but in the nuclear effects,
which can be estimated by taking any reasonable form for
the internal GPD.
Taking into account that 
\begin{equation}
z - { \xi \over \xi'} = z - [ \tilde z ( 1 + \xi  ) - \xi ]
= z + \xi - { p^+ \over P^+} ( 1 + \xi )
= z + \xi  - { p^+ \over \bar P^+}~,
\end{equation}
Eq. (\ref{spec}) can be written in the form
\begin{eqnarray}
H_{q}^3(x,\xi,\Delta^2) =  
\sum_N \int_x^1 { dz \over z}
h_N^3(z, \xi ,\Delta^2 ) 
H_q^N \left( {x \over z},
{\xi \over z},\Delta^2 \right)~,
\label{main}
\end{eqnarray}
where the off-diagonal light cone momentum distribution
\begin{equation}
h_N^3(z, \xi ,\Delta^2 ) =  
\int d E
\int d \vec p
\, P_N^3(\vec p, \vec p + \vec \Delta) 
\delta \left( z + \xi  - { p^+ \over \bar P^+ } \right)
\label{hq0}
\end{equation}
has been introduced.
As it is shown in Ref. \cite{io}, 
Eqs. (\ref{main}) and (\ref{hq0}) or, which is the same,
Eq. (\ref{spec}), fulfill the constraint $i)-iii)$ previously listed.
The constraint $i)$, i.e. the forward limit
of GPDs, is verified
by taking
the forward limit ($\Delta^2 \rightarrow 0, \xi \rightarrow 0$)
of Eq. (\ref{main}), 
yielding the parton distribution $q_3(x)$ in IA:
\cite{fs,cio,io1}: 
\begin{eqnarray}
q_3(x) =  H_q^3(x,0,0) =
\sum_{N} \int_x^1 { dz \over z}
f_{N}^3(z) \,
q_{N}\left( {x \over z}\right)~.
\label{mainf}
\end{eqnarray}
In the latter equation,
\begin{equation}
f_{N}^3(z) = h_{N}^3(z, 0 ,0) =  \int d E \int d \vec p
\, P_{N}^3(\vec p,E) 
\delta\left( z - { p^+ \over \bar P^+ } \right)
\label{hq0f}
\end{equation}
is the forward limit of Eq. (\ref{hq0}), i.e.
the light cone momentum distribution of the nucleon $N$
in the nucleus, $q_N(x)= H_q^N( x , 0, 0)$
is the distribution
of the quark of flavour $q$ 
in the nucleon $N$ and $P_N^3(\vec p, E)$,
the $\Delta^2 \longrightarrow 0$ limit of
Eq. (\ref{main}), is the
one body spectral function.

The constraint $ii)$, i.e. the $x-$integral of the GPD
$H_q$, is also 
fulfilled. By $x-$integrating Eq. (\ref{main}),
one obtains:
\begin{eqnarray}
\int dx H_q^3(x,\xi,\Delta^2) =
\sum_N 
F_q^N(\Delta^2)
F_N^3(\Delta^2)
= F_q^3(\Delta^2)~.
\label{ffc}
\eq
In the equation above,
$F_q^3(\Delta^2)$ is the
contribution, 
of the quark of flavour $q$,
to the
nuclear f.f.;
$F_q^N(\Delta^2)$ is the contribution,
of the quark of flavour $q$,  
to the nucleon $N$ f.f.;
$F_N^3(\Delta^2)$ is the 
so-called $^3He$ ``point like f.f.'', which
would represent the contribution of the nucleon $N$ to the
f.f. of $^3He$ if $N$ were point-like.
$F_N^3(\Delta^2)$ is given, in the present approximation, by
\bq
F_N^3(\Delta^2) = \int dE \int d \vec p
\, P_N^3(\vec p, \vec p + \vec \Delta, E)
= \int dz \, h_N^3(z,\xi,\Delta^2)~. 
\label{ffp}
\eq
Eventually the polynomiality, condition $iii)$,
is formally fulfilled by Eq. (\ref{spec}).

In the following,
$H_q^3(x,\xi,\Delta^2)$, Eq. (\ref{spec}), 
will be evaluated in the nuclear Breit Frame.
The non-diagonal spectral function
Eq. (\ref{spectral}), appearing in Eq.
(\ref{spec}),
will be calculated 
along the lines of Ref. \cite{gema},
by means of 
the overlap Eq. (\ref{trueover}), which 
exactly includes
the final state interactions in the two nucleon recoiling system
\cite{over}. 
The realistic wave functions $\Psi_3^M$
and $\Psi_R^{S_R}$ in Eq. (\ref{trueover})
have been evaluated
using the 
AV18 interaction \cite{av18}.
In particular $\Psi_3^M$ has been 
developed along the lines of Ref. \cite{tre}.
The same overlaps have been already used in Ref.\cite{io1,io}.

The other ingredient in Eq. (\ref{spec}), i.e.
the nucleon GPD $H_q^N$, has been modelled in agreement with
the Double Distribution representation \cite{radd}, as described
in \cite{rad1}:
\begin{eqnarray}
H_q^N(x,\xi,\Delta^2) = \int_{-1}^1 d\tilde x
\int_{-1 + |\tilde x|}^{1-|\tilde x|} 
\delta(\tilde x + \xi  \alpha - x)
\tilde \Phi_{q} (\tilde x, \alpha,\Delta^2) d \alpha~,
\label{hdd}
\end{eqnarray}
using the factorized
ansatz: 
\begin{equation}
\tilde \Phi_{q} (\tilde x, \alpha,\Delta^2) =
h_{q} (\tilde x, \alpha)
\Phi_{q} (\tilde x) 
F_{q}(\Delta^2)~.
\label{ans}
\end{equation}
The expressions for the functions $h_{q} (\tilde x, \alpha)$
and $\Phi_{q} (\tilde x)$ can be found in \cite{io}; I recall here that
the $F_q(\Delta^2)$ term in Eq. (\ref{ans}), i.e. the contribution
of the quark of flavour $q$
to the nucleon form factor, has been obtained from
the experimental values of the proton, $F_1^p$, and
of the neutron, $F_1^n$, Dirac form factors. For the
$u$ and $d$ flavors, neglecting the effect of the 
strange quarks, one has
\bq
F_u (\Delta^2)& = & {1 \over 2} [2 F_1^p(\Delta^2) + F_1^n(\Delta^2)]~,
\nonumber
\\
F_d (\Delta^2)& = & 2 F_1^n(\Delta^2) + F_1^p(\Delta^2)~.
\label{fq}
\eq
The contributions of the flavors $u$ and $d$
to the proton and neutron f.f. are therefore
\bq
F_u^p (\Delta^2)& = & {4 \over 3} F_u(\Delta^2)~,
\nonumber
\\
F_d^p & = & - {1 \over 3} F_d(\Delta^2)~, 
\label{fpq}
\eq
and
\bq
F_u^n (\Delta^2)& = & 
{2 \over 3} F_d(\Delta^2)~,
\nonumber
\\
F_d^n (\Delta^2) & = & - {2 \over 3} F_u(\Delta^2)~,
\label{fnq}
\eq 
respectively.

For the numerical calculations,
use has been made of the parametrization of the nucleon
Dirac f.f. given in Ref. \cite{gari}.
I stress again that the main point of the present study
is not to produce realistic estimates for observables,
but to investigate and discuss nuclear effects,
which do not depend on the form of any well-behaved internal
GPD, whose general structure is safely simulated by Eqs.
(\ref{hdd}) -- (\ref{ans}).
In Ref. \cite{io} it has been shown that the described formalism reproduces 
well, in the proper limits, the IA results for nuclear parton distributions
and form factor. In particular, in the latter case, the IA calculation
reproduces well
the data up to a momentum transfer $-\Delta^2=0.25$ GeV$^2$,
which is enough for the aim of this calculation.
In fact, the region of higher momentum transfer 
is not considered here, being
phenomenologically not relevant for the calculation
of GPDs entering coherent processes.

\section{Discussion of conventional nuclear effects}

In this section, some conventional nuclear effects on the GPDs
of $^3He$ will be discussed. The aim is that of avoiding to mistake
them for exotic ones in possible measurements of nuclear GPDs,
and to stress the relevance of experiments using $^3He$ targets

As already done in Ref. \cite{io},
the full result for $H_q^3$, Eq. (\ref{spec}),
will be 
compared with a prescription
based on the assumptions
that nuclear effects are neglected
and the global $\Delta^2$ dependence is
described 
by
the f.f. of $^3He$:
\bq
H_q^{3,(0)}(x,\xi,\Delta^2) 
= 2 H_q^{3,p}(x,\xi,\Delta^2) + H_q^{3,n}(x,\xi,\Delta^2)~,
\label{app0}
\eq
where the quantity
\bq
H_q^{3,N}(x,\xi,\Delta^2)=  
\tilde H_q^N(x,\xi)
F_q^3 (\Delta^2)
\label{barh}
\eq
represents effectively the flavor $q$ GPD of the bound nucleon 
$N=n,p$ in $^3He$. Its $x$ and $\xi$ dependences, given by 
$\tilde H_q^N(x,\xi)$, 
are the same of the GPD of the free nucleon $N$ (represented 
by Eq. (\ref{hdd})),
while its $\Delta^2$ dependence is governed by the
contribution of the flavor $q$ to the
$^3He$ f.f., $F_q^3(\Delta^2)$.
The effect of nucleon motion
and binding can be shown through 
the ratio
\be
R_q(x,\xi,\Delta^2) = { H_q^3(x,\xi,\Delta^2) \over H_q^{3,(0)}
(x,\xi,\Delta^2)}~, 
\label{rnew}
\eq
i.e. the ratio
of the full result, Eq. (\ref{spec}),
to the approximation Eq. (\ref{app0}).
The latter is evaluated by means of the nucleon GPDs used
as input in the calculation, and taking
$
F_u^3(\Delta^2) = {10 \over 3} F_{ch}^{3}(\Delta^2)~,
\label{fu3}
$
and 
$
F_d^3(\Delta^2) = -{4 \over 3} F_{ch}^{3}(\Delta^2)~,
\label{fd3}
$
where $F^3_{ch}(\Delta^2)$ is the f.f. which is calculated
within the present approach, by means of Eq. (\ref{ffc}).
The coefficients $10/3$ and $-4/3$ are chosen
assuming that the contribution of the
valence quarks of a given flavour to the f.f. of $^3He$
is proportional to their charge. 
The ratio Eq. (\ref{rnew})
shows nuclear effects in a very natural way.
As a matter of facts, its forward limit 
yields an
EMC-like ratio for the parton distribution $q$ and,
if $^3He$ were made of free nucleon at rest, it would be one.
This latter fact can be realized by
observing that the prescription Eq. (\ref{app0})
is obtained by
placing $z=1$, i.e. 
no convolution, into Eq. (\ref{spec}). 
One should note that
the prescription suggested in Ref. \cite{km}
for finite nuclei,
assuming that
the nucleus is a system of almost free
nucleons with approximately the same momenta, 
has the same $\Delta^2$ dependence of the 
prescription Eq. (\ref{app0}).

In Figs. 1 to 7, results will presented concerning: 
A) flavor dependence of nuclear effects; B) binding effects; 
C) dependence on the nucleon-nucleon potential.
\subsection{Flavor dependence of nuclear effects}
In the upper panel of
Fig. 1, the ratio Eq. (\ref{rnew}) is shown for the $u$ and $d$
flavor, in the forward limit,
as a function of $x_3=3x$. The trend is clearly EMC-like.
It is seen that nuclear effects for the $d$ flavour are very slightly bigger
than those for the $u$ flavour.
The reason is understood thinking that, in the forward limit,
the nuclear effects are governed by the light cone momentum distribution,
Eq. (\ref{hq0f}): no effects would be
found if such a function were a delta function, while effects get bigger
and bigger if its width increases.
In the lower panel of the same figure,
the light cone momentum distribution, Eq. (\ref{hq0f}),
for the proton (neutron) in $^3He$ is represented by the 
dashed  (full) line.
The neutron distribution is slightly wider than 
the proton one, meaning that the average momentum of the neutron
in $^3He$ is a little larger than the one of the proton
\cite{over,cio,plc}. 
Since the forward $d$ distribution
is more sensitive than the $u$ one to the neutron
light cone momentum distribution, nuclear effects for $d$ are slightly
larger than for $u$, as seen in the upper panel of the same figure.

In Fig. 2, the same analysis of Fig. 1 is performed,
but at $\Delta^2=-0.25$ GeV$^2$ and $\xi_3=3\xi=0.2$.
In this case, nuclear effects are governed by the non-diagonal
light cone momentum distribution, Eq. (\ref{hq0}), shown 
in the lower panel of the figure. In this case, the difference
between the neutron and proton distributions is quite bigger
than in the forward case, governing the difference 
in the ratio Eq. (\ref{rnew}) for the two flavors, which is of the order
of 10 \%, as it is seen in Fig. 3.

From Figs. 1-3 three main conclusions can be drawn.
1) First of all, if one infers properties of nuclear GPDs thinking
to those of nuclear PDs, conventional nuclear
effects as big as 10 \% can be easily
lost, or mistaken for exotic ones.
2) Secondly, this behavior is a typical conventional effect, being
a prediction of IA in DIS off nuclei. If a 10 \% effect would
be observable in experimental studies of nuclear GPDs,
the presence of such a flavor dependence, or its absence,
would be clear signatures of the reaction mechanism of DIS
off nuclei. Its presence would mean that the reaction involves 
essentially partons
inside nucleons, whose dynamics is governed by a realistic
potential in a conventional scenario; 
on the contrary, its absence would mean that,
in a different, exotic scenario,
other degrees of freedom have to be advocated. 
3) Eventually, it is clear that, for this kind of studies, $^3He$
is a unique target, for which experiments are worth to be done:
the flavor dependence cannot be investigated with
isoscalar targets, such as $^2H$ or $^4He$, while for heavier nuclei
calculations cannot be performed with comparable precision.
\subsection{Binding effects}
In the previous section it has been explained how
Eq. (\ref{main}) takes
into account properly the nucleon momentum and energy distributions
through a non-diagonal spectral function.
In the following,
the performances will be studied of three
approximations with increasing complexity
of the full result, defined considering:
i) binding and momentum effects simulated by a
rescaling of variables;
ii) binding effects neglected by using a momentum distribution;
iii) average binding effects evaluated within a simple model
of the spectral function.

i) Binding and momentum effects simulated by a rescaling of variables.

A prescription for nuclear GPDs has been proposed, based on a rescaling
of the variables \cite{km}:
\bq
H_{u}^{3,(1)}(x,\xi,\Delta^2) = F_u^3(\Delta^2) 
\left |
{ d x_N \over d x }
\right |
\theta(|x_N| \leq 1) [ Z H^u(x_N,\xi_N,0) + N H^d(x_N,\xi_N,0)]
\label{app1}
\eq
for the $u$ flavour; the analogous
expression for the $d$ flavour is obtained
by isospin symmetry.
In the above equation, one has, for $^3He$
\be
x_N = 3x {1+\xi_N \over 1 + \xi}~,
\eq
and 
\be
\xi_N = {3 \xi \over (A-1)\xi -1 }~,
\eq
while $F_u^3(\Delta^2)$ is the flavour $u$ contribution to the
$^3He$ f.f., to be fixed by experimental data.
The reliability of the approximation Eq. (\ref{app1}) can be
established studying the ratio: 
\bq
R_q^{(1)} (x,\xi,\Delta^2) = {H_q^3(x,\xi,\Delta^2) 
\over H_{q}^{3,(1)}(x,\xi,\Delta^2)}~,
\label{rat2}
\eq
where the numerator is given by the full result,
Eq. (\ref{main}), and the denominator by Eq. (\ref{app1}).
This comparison permits to estimate to what extent 
Eq. (\ref{app1}) describes effectively 
the nucleon motion and binding.
Such a prescription has been indeed used to
parametrize nuclear GPDs for estimates
of DVCS cross sections and asymmetries for finite nuclei.
To have a consistent check,
the denominator in Eq. (\ref{rat2}), i.e. Eq.
(\ref{app1}), has been evaluated through the model
for $H_q^N$ used in the full calculation, together
with the $u$ contribution to the $^3He$ f.f.
previously used.
Results are presented in Fig. 4, where
the ratio Eq. (\ref{rat2}) is shown for the $d$ flavor,
in the forward limit and 
for $\Delta^2 = -0.25$ GeV$^2$ and $\xi_3=0.2$.
It is clearly seen that, while the approximation
Eq. (\ref{app1}) differs from the full result
by at most 5 \% in the forward limit,
yielding something similar to an EMC-like effect,
it differs systematically by more than 10 \% for all the values of $x_3$. 
In general, the prescription Eq. (\ref{app1}) does not
describe effectively the conventional nuclear effects,
not even in a rough way, and one should not use it, at least
for light nuclei. 

ii) Binding effects neglected by using a momentum distribution.

In the previous section, it has been shown that IA leads
to nuclear effects governed by a one body non-diagonal spectral function.
This means that overlap integrals involving excited states
with a given excitation energy, $E_R^*$, of the
two body recoiling system, have to be evaluated.
When a momentum distribution is used instead of a spectral
function, not only the IA, but also another approximation,
the so called ``closure approximation'', has been used:
an average excitation energy, $\bar E^*$, has been inserted
in the expression of the delta function appearing in the definition
of the spectral function Eq. (\ref{spectral}), so that the completeness
of the two body recoiling states can be used \cite{cio}:
\be
P_N^3(\vec p, \vec p + \vec \Delta, E)  & \simeq &
\, \bar{\sum}_{M} 
\sum_{s}
\bra \vec P'M | a_{\vec p + \vec \Delta,s}
a^\dagger_{\vec p, s}| \vec P M \ket
\delta(E - E_{min} - \bar E^*)
\nonumber
\\
& = &
n(\vec p, \vec p + \vec \Delta)\,\delta(E - E_{min} - \bar E^*)~,
\label{clos}
\eq
and the spectral function is approximated by a
one-body non diagonal momentum distribution times
a delta function defining an average 
value of the removal energy.
Whenever the momentum distribution is used instead of the spectral
function, in addition to the IA the above closure approximation has been
used assuming  $\bar E^*=0$, i.e., binding effects have been completely
neglected.
The difference between the full calculation and the
one using the momentum distribution, for the ratio
Eq. (\ref{rnew}), is shown in Fig. 5.
It is seen that, while the difference is a few percent
in the forward limit, it grows in the non-forward case,
becoming an effect of 5 \% to 10 \% between 
$x = 0.4$ and $0.7$.
From this analysis one should draw three main conclusions,
basically the same arisen in the study of the flavor dependence:
1) the size of nuclear effects found for GPDs is bigger
than that found in inclusive DIS;
2) if the conventional binding were not taken into account, 
a possible 10 \% effect, found experimentally, could be mistaken 
for an exotic effect;
3) $^3He$ is a unique target to study the binding effects, since the
realistic evaluation of non diagonal spectral functions 
is a challenging task for heavier targets. 

iii) Average binding effects evaluated within 
a simple model of the spectral function.

A little more refined calculation can be performed using, in Eq. (\ref{clos}),
an average value of the removal energy, instead of the minimum one.
A calculation has been performed assuming, as average values of the 
excitation energy, the values calculated by means of the spectral functions 
corresponding to the AV18 interaction. The obtained values are in agreement
with the ones listed in Ref. \cite{plc}.
In Fig. 6 it is shown that the situation, if compared
with the previous calculation, performed using a 
momentum distribution only, improves a little.
In any case, the difference becomes negligible in the forward
limit, in agreement with the findings of Ref. \cite{cio}, while it keeps
being sizable in the non forward one.
One should also keep in mind that, in general, for $^3He$, which is
a loosely bound system, binding effects are smaller than for
heavier nuclei. Moreover, for the latter systems, for which
realistic spectral functions are not available, the evaluation
of the average removal energy is not easy and it is 
affected by theoretical uncertainties.
Therefore the conclusions
of the previous subsection, concerning the importance and relevance
of binding effects in studies of nuclear GPDs, are confirmed.
\subsection{Dependence on the nucleon-nucleon potential}
In Fig. 7, the difference is shown between the full calculation,
Eq. (\ref{rnew}), evaluated with the AV18 interaction \cite{av18}, 
and the same quantity, evaluated by means of the AV14 one.
It is seen that there is basically no difference in the forward
limit, confirming previous findings in inclusive DIS
\cite{io1}, while a sizable difference is seen in the non forward case
(preliminary results of this behavior have been accounted for in a talk
at a Conference \cite{io}).
From these analyses the same conclusions of the previous two subsections
can be drawn:
1) properties of nuclear GPDs should not be naively inferred by
those of nuclear PDs;
2) if conventional effects were not properly evaluated 
they could be mistaken 
for exotic ones in the analysis of the data;
3) $^3He$, for which the best realistic calculations
are possible, is a unique target to study these effects.
We note on passing that a difference between observables evaluated
using AV18 and AV14 potentials is not easily found, in particular
in inclusive DIS.

\section{GPDs for the $^3H$ target}

In the perspective of using $^3H$ targets after the 12 GeV upgrade
of JLab \cite{wp}, it is useful to address what could be learnt 
from simultaneous measurements with trinucleon targets, $^3He$ and $^3H$.

The procedure proposed firstly in Ref. \cite{thom1} for the unpolarized
DIS to extract, with unprecedented precision,
the ratio of down to up quarks in the proton, $d(x)/u(x)$,
at large Bjorken $x$, 
is extended here to the case of the GPDs of trinucleons.
To minimize nuclear effects, 
the following ``super-ratio'', a generalization
of the one proposed in Ref. \cite{thom1},  can be defined
\be
S_{qq'}(x,\xi,\Delta^2) = R_q^H(x,\xi,\Delta^2)/R_{q'}^T(x,\xi,\Delta^2)~,
\label{sr}
\eq
where the ratio
\be
R_q^A(x,\xi,\Delta^2)= { H_q^A(x,\xi,\Delta^2)
\over
Z_A H_q^p(x,\xi,\Delta^2) + N_A H_q^n(x,\xi,\Delta^2) }~,
\label{rsr}
\eq
has been introduced for $^3He$ ($A=H$) and $^3H$ ($A=T$),
with $q=u,d$, $Z_A (N_A)$ the number of protons (neutrons)
in the nucleus $A$, and $H_q^N(x,\xi,\Delta^2)$ the GPD
of the quark $q$ in the nucleon $N=p,n$.
Now, using the isospin symmetry of GPDs \cite{dpr},
we can call
\begin{eqnarray}
H_u(x,\xi,\Delta^2) & = &
H_u^p(x,\xi,\Delta^2)
=H_d^n(x,\xi,\Delta^2)~,
\\
H_d(x,\xi,\Delta^2) & = &
H_d^p(x,\xi,\Delta^2)
=H_u^n(x,\xi,\Delta^2)~,
\end{eqnarray}
so that Eq. (\ref{sr}) is given, for example for $q=d$ and $q'=u$,
by the simple relation
\bq
S_{du} (x,\xi,\Delta^2)= { H_d^H(x,\xi,\Delta^2) \over H_u^T(x,\xi,\Delta^2)}~,
\label{extr}
\eq
a quantity in principle observable.

In the IA approach discussed here, using Eq. (\ref{main})
to calculate the nuclear GPDs, one has therefore
\bq
S_{du} (x,\xi,\Delta^2)= { 
\int_x^1 {dz \over z} \left \{ 
h_p^H(z,\xi,\Delta^2) H_d \left({x \over z},{\xi \over z },\Delta^2 \right )
+
h_n^H(z,\xi,\Delta^2) H_u \left({x \over z},{\xi \over z },\Delta^2 \right )
\right \}
\over 
\int_x^1 {dz \over z} \left \{ 
h_n^T(z,\xi,\Delta^2) H_d \left({x \over z},{\xi \over z },\Delta^2 \right )
+
h_p^T(z,\xi,\Delta^2) H_u \left({x \over z},{\xi \over z },\Delta^2 \right )
\right \} }~,
\label{extr}
\eq
where $h_{p(n)}^{H(T)}(z,\xi,\Delta^2)$ represents 
the light cone off diagonal momentum distribution for
the proton (neutron) in $^3He$ ($^3H$).
If the Isospin Symmetry were valid at the nuclear level,
one should have 
$h_{p}^{H}(z,\xi,\Delta^2) = h_{n}^{T}(z,\xi,\Delta^2)$, 
and
$h_{n}^{H}(z,\xi,\Delta^2) = h_{p}^{T}(z,\xi,\Delta^2)$,
so that the ratio Eq. (\ref{extr}) would be identically 1.
From the analysis of Section III and the Figures 1 and 2
it is clear anyway
that these relations are only approximately true,
and some deviations are expected.

In Fig. 8, the super-ratio $S_{du} (x,\xi,\Delta^2)$, Eq. (\ref{sr}),
evaluated by using the AV18 interaction for the nuclear GPDs
in Eq. (\ref{main}),
taking into account therefore the Coulomb interaction between
the protons in $^3He$ and a weak charge independence breaking term, 
is shown for different values of $\Delta^2 \leq 0.25$ GeV$^2$ and $\xi$.
While it is seen that, as expected, $S_{du} (x,\xi,\Delta^2)$
is not exactly 1 and the difference gets bigger with increasing
$\Delta^2$ and $\xi$, for the low values of $\Delta^2$ and $\xi$ relevant
for the present investigation of GPDs, 
such a difference keeps being a few percent one.

It would be very interesting to measure this ratio experimentally.
If strong deviations from this predicted behavior were observed,
there would be a clear evidence that the description in terms of
IA, i.e. in terms of the conventional scenario of partons confined
in nucleons bound together by a realistic interaction, breaks down.
In other words one could have a clear signature of possible interesting
exotic effects.
One should notice that the trend obtained for the ratio 
Eq. (\ref{sr}) is almost flat;
this may have to do with the simple model used for the nucleon GPD
in the convolution formula.
The present analysis permits
only to estimate the size of nuclear effects between the forward
and the general case, and not to predict the $x$ behavior of the ratio with 
good precision. In any case,
the possibilities offered by a $^3H$ target deserve more attention
and will be discussed in more detail elsewhere.

\section{Conclusions}

In this paper, using
a realistic microscopic calculation, 
some peculiar conventional nuclear effects on
the unpolarized quark GPD $H_q^3(x,\xi,\Delta^2)$ for $^3He$
have been investigated. 
By studying the dependence of nuclear effects on the flavor, the
nucleon binding and the nucleon-nucleon potential, the same three main
conclusions have been obtained:
1) the size of nuclear effects found in inclusive DIS
should not be naively used to estimate the one expected for GPDs;
2) if conventional nuclear physics is not taken into account properly, 
several experimentally observable effects, each of the order of 10 \%, 
could be mistaken for exotic ones;
3) trinucleons represent  unique targets to study these effects, since 
a comparably realistic treatment for heavier targets
is a challenging task. 
It has also been shown that the simultaneous use of a $^3He$
and $^3H$ target would help disentangling the conventional effects
from the exotic ones.

The issue of applying the obtained GPDs to  
estimate cross-sections and to establish
the feasibility of experiments, is in progress
and will be presented elsewhere.
In particular, the study of polarized GPDs
will be very interesting, due to the peculiar
spin structure of $^3He$ and its implications
for the study of the angular momentum of the free neutron.

\acknowledgments

I would like to thank the organizers of the meeting
``Journ\'ees Noyaux du GDR Nucl\'eon'', 
held in Grenoble
in November 2008,
for the invitation and the useful and pleasant discussions.

\newpage

\section*{Figure Captions}

\vspace{1em}\noindent
{\bf Fig. 1}:
Upper panel: the dashed (full) line represents
the ratio Eq. (\ref{rnew}), for the $u$ ($d$)
flavor, in the forward limit.
Lower panel: the dashed (full) line represents
the light cone momentum distribution, Eq. (\ref{hq0f}),
for the proton (neutron) in $^3He$.

\vspace{1em}\noindent
{\bf Fig. 2}:
Upper panel: the dashed (full) line represents
the ratio Eq. (\ref{rnew}), for the $u$ ($d$)
flavor, at $\Delta^2=-0.25$ GeV$^2$ and $\xi_3=0.2$.
Lower panel: the dashed (full) line
represents the light cone off diagonal momentum distribution, 
Eq. (\ref{hq0}),
for the proton (neutron) in $^3He$,
at  $\Delta^2=-0.25$ GeV$^2$ and $\xi_3=0.2$.

\vspace{1em}\noindent
{\bf Fig. 3}:
The ratio of the ratios Eq. (\ref{rnew}), for the $d$ to the
$u$ flavor, at  $\Delta^2=-0.25$ GeV$^2$ and $\xi_3=0.2$ (full line)
and in the forward limit (dashed line).
 
\vspace{1em}\noindent
{\bf Fig. 4}: 
The ratio Eq. (\ref{rat2}), for the $d$ flavor, in the forward limit
(full line) and at  $\Delta^2=-0.25$ GeV$^2$ and $\xi_3=0.2$ (dashed line).

\vspace{1em}\noindent
{\bf Fig. 5}:
Upper panel: the ratio Eq. (\ref{rnew}), in the forward limit,
for the $d$ flavor,
corresponding to the full result of the present approach
(full line), compared with the one obtained using in the
numerator the approximation Eq. (\ref{clos}) with $\bar E^*=0$,
i.e., using a momentum distribution instead of a spectral function
(dashed line).
Lower panel: the same as before, but 
evaluated at $\Delta^2=-0.25$ GeV$^2$ and $\xi_3=0.2$.  

\vspace{1em}\noindent
{\bf Fig. 6}:
The same of Fig. 5, but using the approximation Eq. (\ref{clos})
with the values of $\bar E^*$ obtained from the spectral function
corresponding to the AV18 interaction. 

\vspace{1em}\noindent
{\bf Fig. 7}:
Upper panel: the ratio Eq. (\ref{rnew}), in the forward
limit, for the $d$ flavor,
corresponding to the full result of the present approach,
where use is made of the AV18 interaction
(full line), compared with the one obtained using in the
numerator the AV14 interaction (dashed line): the two curves
cannot be distinguished.
Lower panel: the same, but 
evaluated $\Delta^2=-0.25$ GeV$^2$ and $\xi_3=0.2$: now the curves are
distinguishable.  

\vspace{1em}\noindent
{\bf Fig. 8}:
The ratio Eq. (\ref{extr}), in the forward limit (dot-dashed line),
at $\Delta^2=-0.15$ GeV$^2$ and $\xi_3=0.1$ (dashed line),
and at
$\Delta^2=-0.25$ GeV$^2$ and $\xi_3=0.2$ (full line).

\newpage

\begin{figure}[h]
\vspace{13.0cm}
\includegraphics{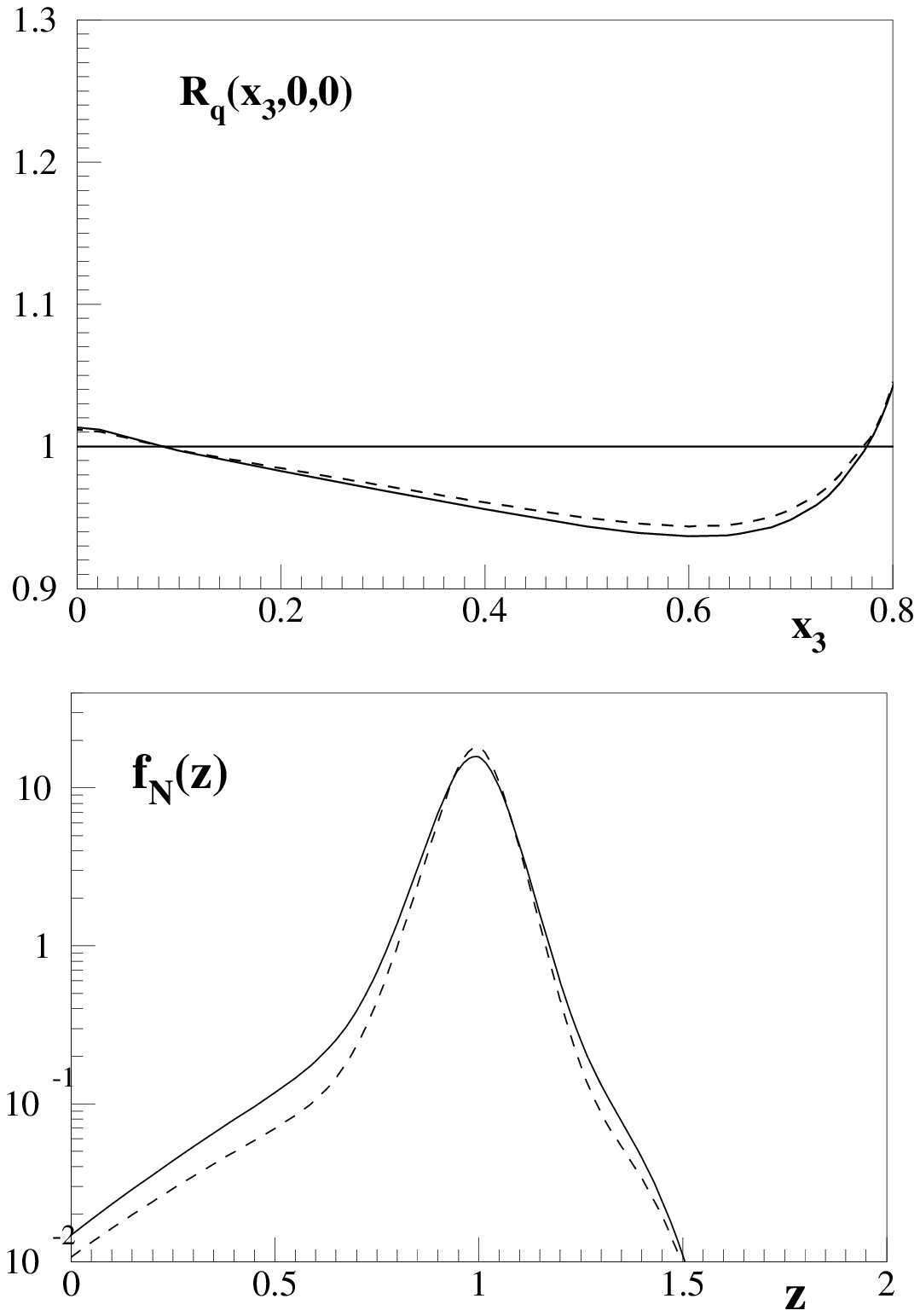}
\caption{}
\end{figure}

\newpage
$ $
\begin{figure}[h]
\vspace{13.0cm}
\includegraphics{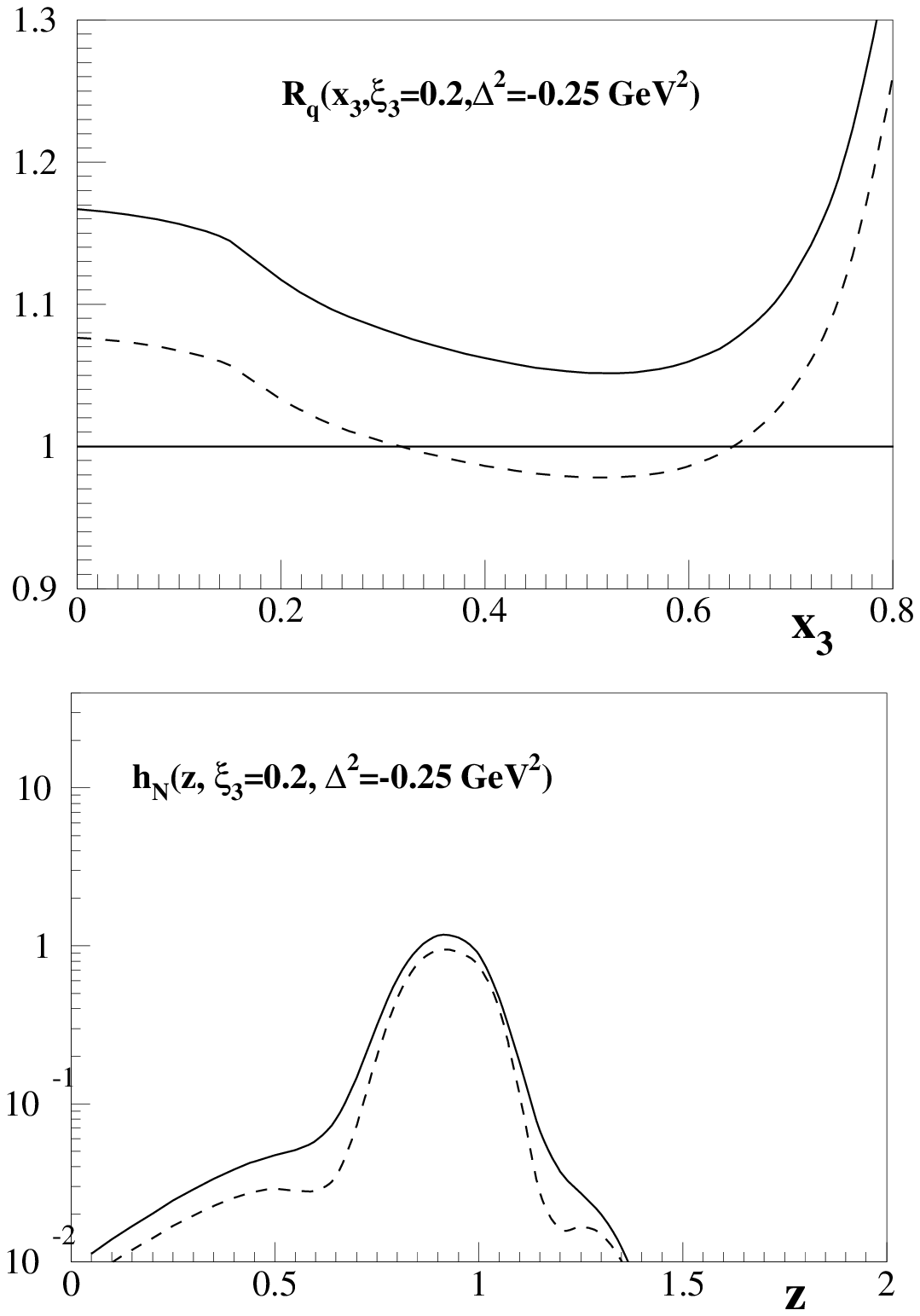}
\caption{}
\end{figure}

\newpage
$ $
\begin{figure}
\vspace{10.0cm}
\includegraphics{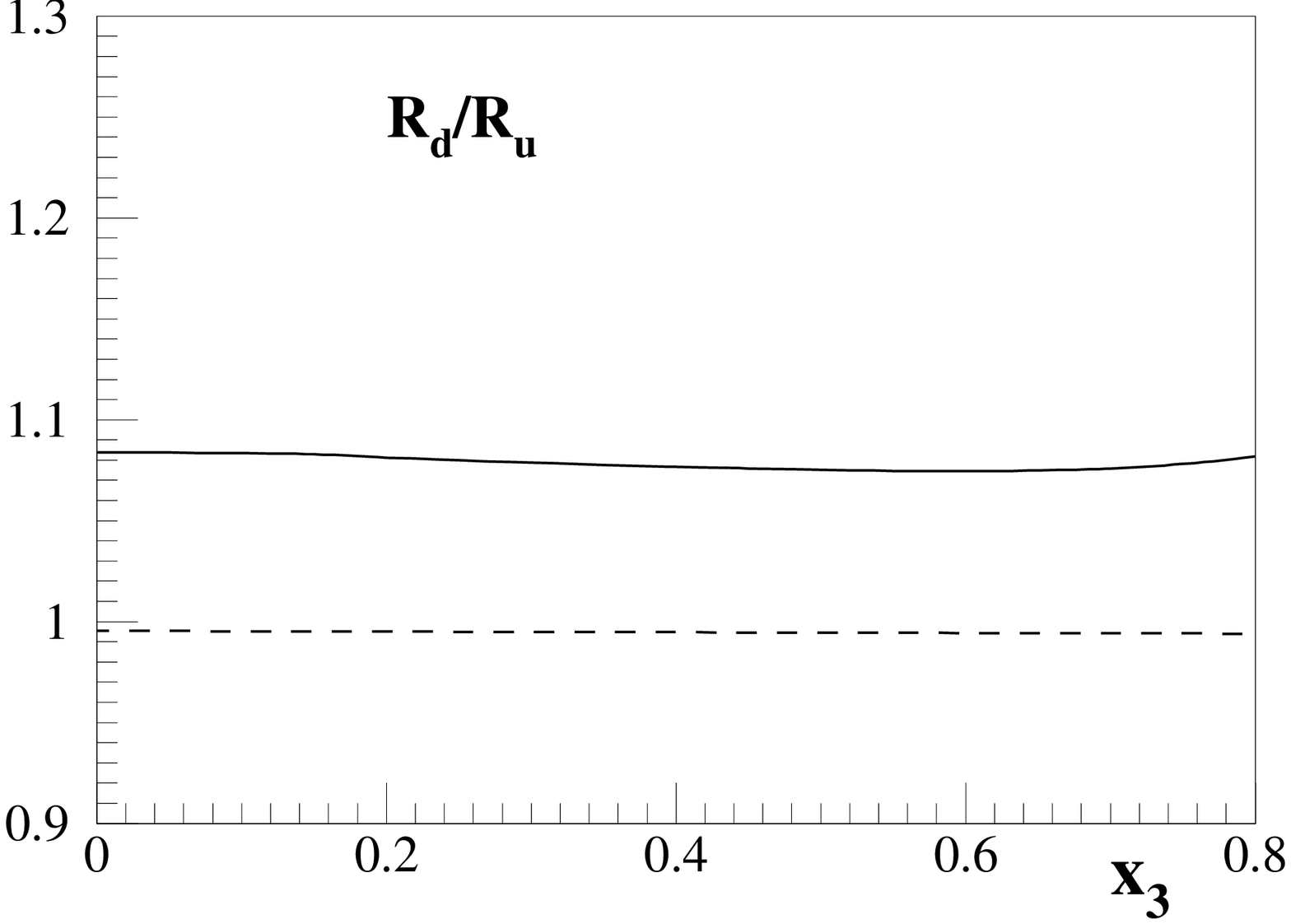}
\caption{}
\end{figure}

\newpage
$ $
\begin{figure}
\vspace{10.0cm}
\includegraphics{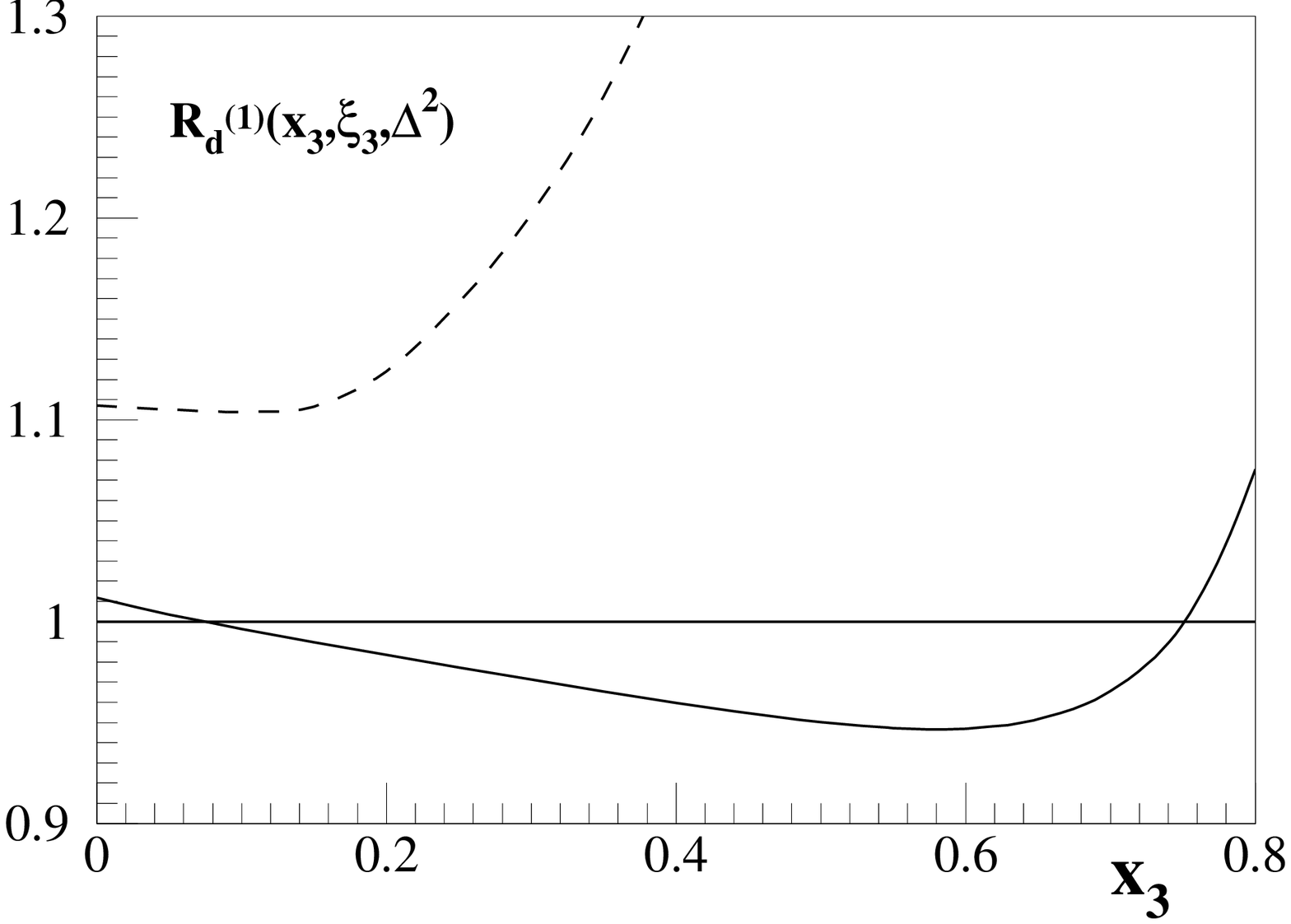}
\caption{}
\end{figure}

\newpage
$ $
\begin{figure}[h]
\vspace{13cm}
\includegraphics{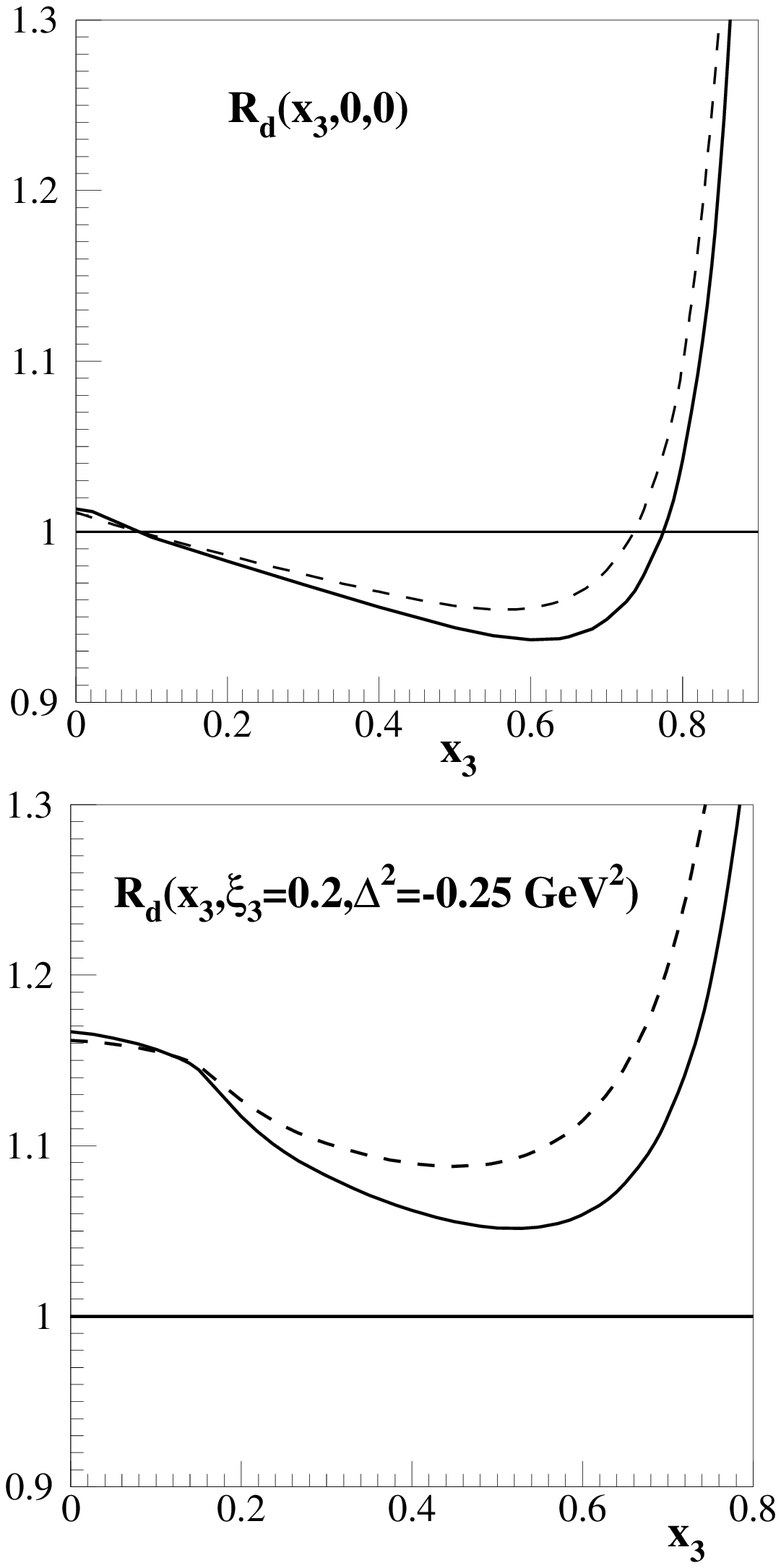}
\caption{}
\end{figure}

\newpage
$ $
\begin{figure}[h]
\vspace{13.cm}
\includegraphics{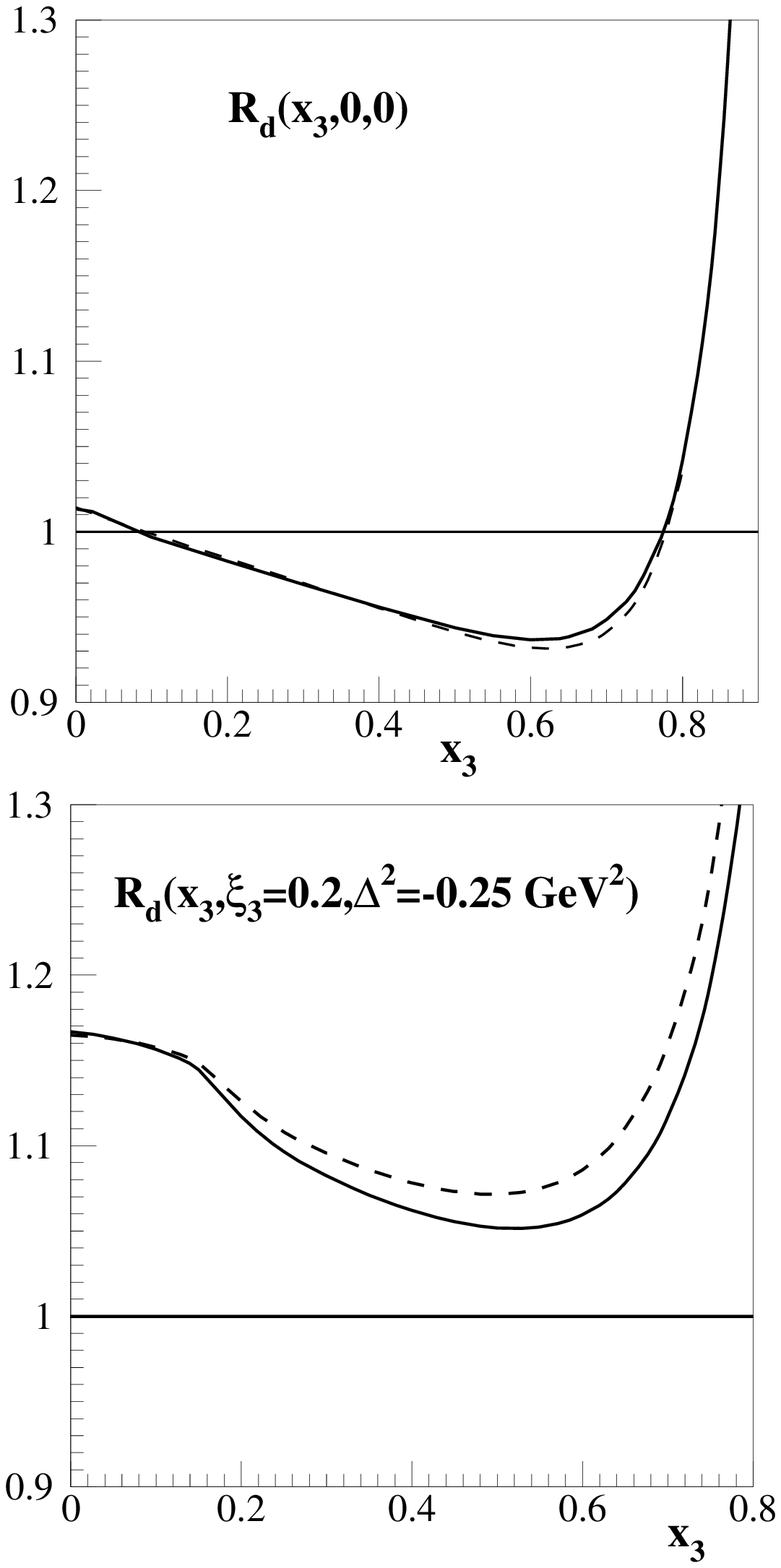}
\caption{}
\end{figure}

\newpage
$ $
\begin{figure}[h]
\vspace{13.cm}
\includegraphics{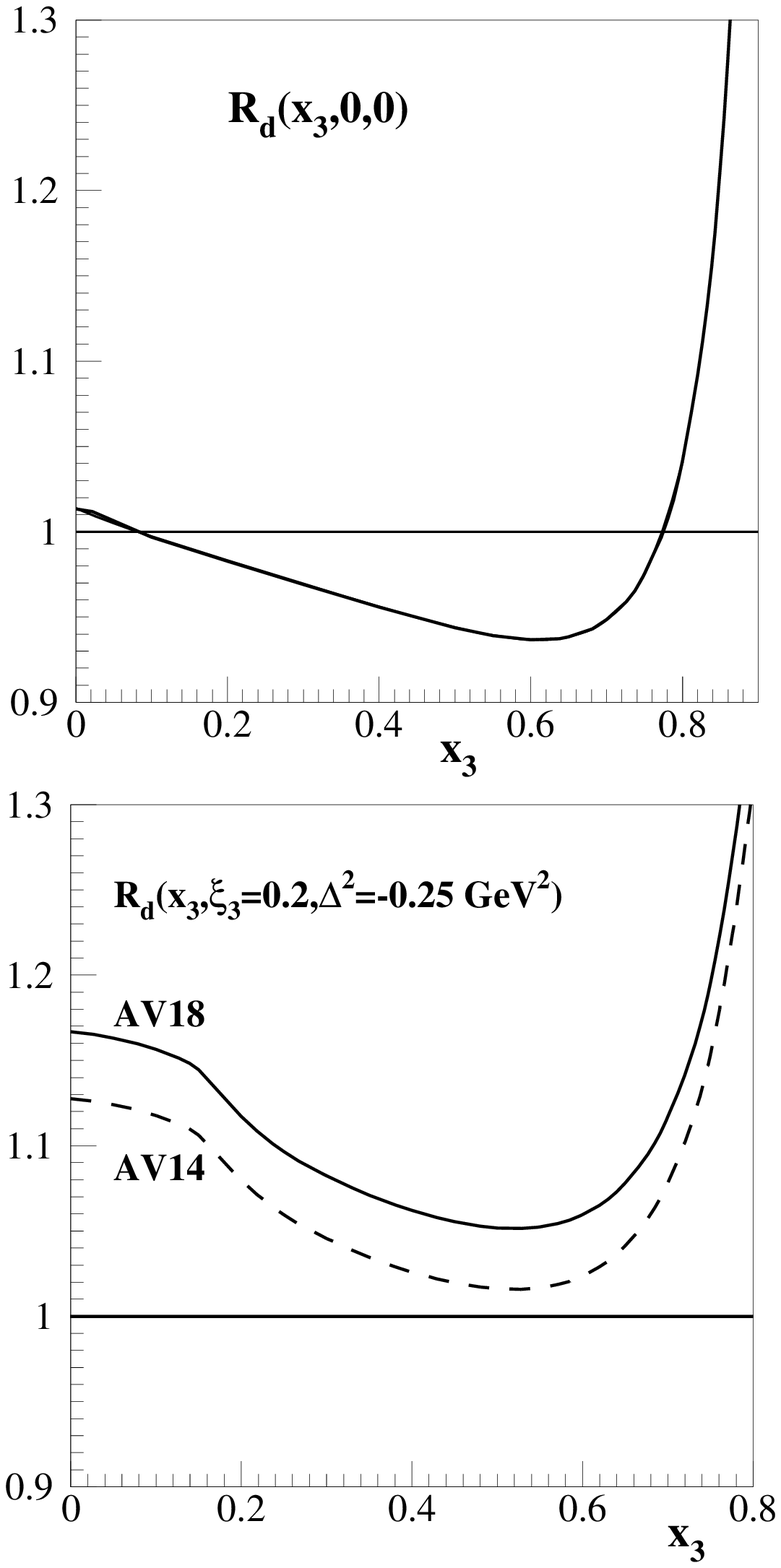}
\caption{}
\end{figure}

\newpage
$ $
\begin{figure}
\vspace{10.0cm}
\includegraphics{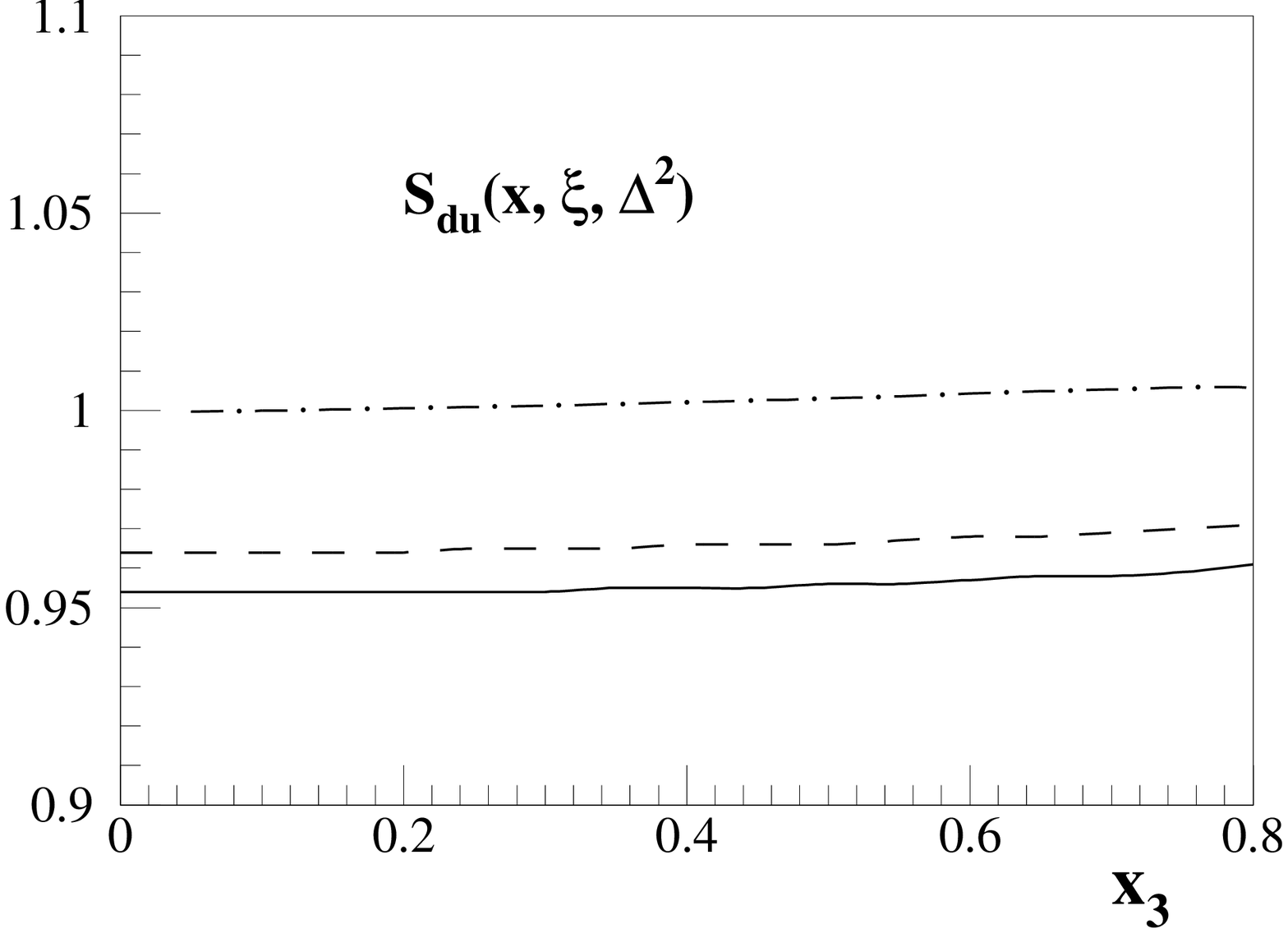}
\caption{}
\end{figure}

\end{document}